\documentclass[epj,draft]{special}
\usepackage{psfig}
\begin{document}
\title{Semi-leptonic B-decays and the 
two-pion distribution amplitudes}
\subtitle{LU TP 01-11}
\author{Martin Maul% etc
}                     % Do not remove
\offprints{}          % Insert a name or remove this line
\institute{Department of Theoretical Physics, Lund University,
         S\"olvegatan 14A, S - 223 62 Lund, Sweden}
\date{\today}
%Received: date / Revised version: date}
\abstract{We show that the semi-leptonic decay 
$B^+ \to  \pi^+\pi^-\ell^+\nu_\ell$ can be used as a source of information
for two-pion distribution amplitudes. The connection between these amplitudes
and the B-meson decay width is achieved by the light cone sum rule method. We 
show the relevant distribution amplitudes and give the formula for the decay
width.
 % end of PACS codes
} %end of abstract
\maketitle
\section{Introduction}
\label{intro}
Recently, two-pion distribution amplitudes (2$\pi$DAs) 
have received considerable interest \cite{Polyakov:1999ze,Polyakov:1999td}
because of their relation
to skewed parton distributions \cite{Polyakov:1999gs}.
From an experimental point of view the 2$\pi$DAs have been mostly
discussed for the reaction $\gamma\gamma^* \to \pi\pi$
\cite{Kivel:1999sd,Kivel:2000rq,Maul:2001ky,Diehl:2000uv}, where the 
factorization has been proven in Refs.~\cite{Diehl:1998dk,Freund:2000xg}, and
also in hard exclusive electroproduction 
\cite{Lehmann-Dronke:2000aq,Lehmann-Dronke:1999ym}.
Here, we want to add another type of reaction which could
provide valuable insight in the dynamics of the formation of two pions,
namely semi-leptonic B-decays.
\newline \newline
The reaction $B\to\pi\pi\ell\nu$ is an alternative
 source of information about
the 2$\pi$DAs as compared 
to $\gamma\gamma^*\to \pi\pi$ because new structures 
arise due to the fact that the semi-leptonic weak decay induces an axial 
vector current in addition to the vector current and the fact
that the B-meson is a pseudoscalar particle.

The method used here to connect the 2$\pi$DAs with the 
B-meson decay width is the
method of Light Cone Sum Rules  (LCSR) where the decay 
amplitude $B\to \pi\pi\ell \nu$ is related to the light-cone
OPE of the corresponding  
correlation function where factorization into the 2$\pi$DAs and 
the hard amplitude is guaranteed by the large virtuality of the off-shell
currents. 
The light-cone sum rule method
applied here  is essentially the same 
as used in $B\to\pi\ell \nu$ \cite{Belyaev:1993wp,Khodjamirian:1998ji} 
and $B\to\rho\ell \nu$ \cite{Ball:1997rj}, with the only distinction
that the 2$\pi$DAs  enter. The advance of 
B-factories may yield a lot of new experimental data on the decay 
$B\to\pi\pi\ell\nu$, where explicit models for the various 2$\pi$DAs
entering in this process may be tested and could yield a deepening 
understanding of the non-perturbative multi-particle dynamics which lies
behind these generalized distribution amplitudes.
\section{The method}
\subsection{Kinematics} 
\label{kinem}
We consider the process $B^+ \to  \pi^+\pi^-\ell^+\nu_\ell$. 
The kinematics of the process
is given by $B(q)\to \ell(p_e)+ \nu(q'-p_e) +\pi^+(k_1)+\pi^-(k_2)$ and there 
exist two light-like vectors $n^+$ and $n^-$ with $n^+n^-=1/2$ such that:
\begin{eqnarray}
q &=& (m_B,0,0,0) = m_B\left(n^++n^-\right)
\nonumber \\
q' &=&  {q'}^- n^+ + {q'}^+ n^-
\label{lightlike}
\end{eqnarray}
in the rest frame of the B-meson. We can now make a division into
 'good' (+) and 'bad' (-) components, where the 'bad'
components can be neglected  requiring  
$m_B  \gg q'_- >0$ and $m_B^2  \gg  P^2 = (k_1+k_2)^2 = W^2$. 
This requirement is necessary to ensure the factorization in the 
approach of the LCSR technique at the stage where the virtual
amplitude is factorized in a hard scattering part and the 2$\pi$DAs. 
The factorization follows in complete analogy from the $\gamma \gamma^*$
case.  Under these
circumstances $P^2/ {m_B'}^2$, using $m_B' = m_B -{q'}_-$
can be considered as a small expansion parameter
and we can decompose:
\begin{eqnarray}
P &=& q-q' = k_1+k_2 = m_B'n^+ +  \frac{P^2}{m_B'}n^-
\nonumber \\
k_1 &=& \zeta m_B' n^+ + \bar \zeta \frac{P^2}{m_B'}n^- + K_\perp
\nonumber \\
k_2 &=& \bar \zeta m_B' n^+ +  \zeta \frac{P^2}{m_B'}n^- - K_\perp\;,
\end{eqnarray}
using $\bar \zeta = 1-\zeta$. In this way we have set up a similar light
cone decomposition as in the case $\gamma\gamma^*\to \pi\pi$. 
\newline \newline
For the light cone sum rule technique to apply we need to consider a situation
where  the B meson is  off shell. 
One can achieve this in the frame-work of the kinematics discussed so far by simply changing:
\begin{equation}
q \to m_Bn^+ + \frac{q^2}{m_B}n^-\;.
\end{equation} 
The two light-like vectors $n^+$ and $n^-$ are still the same as in Eq.~(\ref{lightlike}).
\subsection{The correlator and the distribution amplitudes}
\label{distamp}
\begin{figure}
\centerline{\psfig{figure=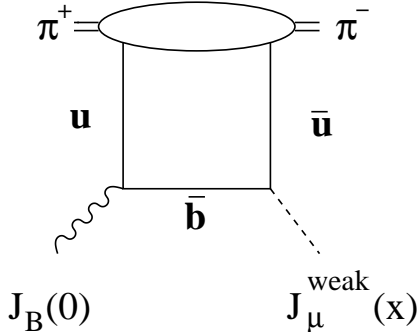,width=5.5cm}}
\caption{Diagrammatic representation of the correlation function. The wavy line
represents the external 'B-meson current' $J_B$ and the dashed line the weak current
$J^{\rm weak}$.}
\label{diagram}
\end{figure}

For the application of the LCSR approach one considers
a correlation function of a pseudoscalar and a weak current with
a quark content that corresponds to the one of a $B^\pm$ meson:
\begin{eqnarray}
T_{\mu}(q,q') &=& 
i \int d^4x e^{-iq'x} \langle 0 | T[J_B(0),J^{\rm weak}_\mu(x)] 
|\pi^+\pi^-(P) \rangle
\nonumber \\
J_B(x) &=& \bar u(x) i\gamma_5 b(x)
\nonumber \\
J^{\rm weak}_\mu(x) &=& 
 \bar b(x)i \gamma_\mu (1-\gamma_5) u(x)\;.
\label{corr}
\end{eqnarray}
In this correlation function the B-meson is interpolated by a current 
$J_B$ with four-momentum $q$, which is off shell. 
$J_\mu^{\rm weak}$ is the weak $b\to u$ transition current.
The diagrammatic representation of $T_\mu$ is shown in Fig.~\ref{diagram}.
Depending on their angular momentum the two pions in the final state
can have odd or even parity. With the definition:
\begin{equation}
\langle \pi^+\pi^- (P)| \bar q(x) \Gamma q(0)|0\rangle_{x^2 = 0}
=\Phi^{[\Gamma]}_q \;,
\end{equation}
we can parameterize the structures that will occur in our correlation
function to leading-twist accuracy: 
\begin{eqnarray}
{\rm twist-2:} && 
\nonumber\\
\Phi^{[i \sigma_{\mu\nu}]}_q 
&=& \frac{\left(k_{1\mu}k_{2\nu} - k_{2\mu}k_{1\nu}\right)}{2W}
 \int_0^1 dz \;f_{\perp q}^{\pi\pi}(z,\zeta, W^2) e^{izP\cdot x}  
\nonumber \\
&=& \frac{\left(P_{\mu}R_{\nu} - R_{\mu}P_{\nu}\right)}{4W}
 \int_0^1 dz \;f_{\perp q}^{\pi\pi}(z,\zeta, W^2) e^{izP\cdot x}  
\nonumber \\
\Phi^{[{\gamma_\mu\gamma_5}]}_q 
&=& \epsilon_{\mu k_1 k_2 x}
 \int_0^1 dz \;g_{q}^{\pi\pi}(z,\zeta, W^2) e^{izP\cdot x}  
\nonumber \\
&=& \epsilon_{\mu P R x}
 \int_0^1 dz \;g_{q}^{\pi\pi}(z,\zeta, W^2) e^{izP\cdot x}  
\nonumber \\
\Phi^{[{ \gamma_\mu}]}_q 
&=& P_\mu \int_0^1 dz \;f_{q}^{\pi\pi}(z,\zeta, W^2) e^{izP\cdot x}  
\nonumber \\
{\rm twist-3:}&& \nonumber \\
\Phi^{[{\bf 1}]}_q &=& W \int_0^1 dz 
\;e_q^{\pi\pi}(z,\zeta, W^2) e^{izP\cdot x} \nonumber \\ 
\Phi^{[{ \gamma_5}]}_q 
&=& 0 \;,
\end{eqnarray} 
using $R_\mu = k_{2\mu} -  k_{1\mu}$. Especially, we find
$P\cdot R = 0$ and $R^2 = 4 m_\pi^2 -  W^2$.
We only need one of the two Lorentz structures $i \sigma^{\mu\nu}\gamma_5$
and $\sigma^{\mu\nu}$, as they are related to each other by:
\begin{equation}
\sigma^{\mu\nu} = \frac{i}{2} \epsilon^{\mu\nu\alpha\beta} \sigma_{\alpha\beta}
\gamma_5\;.
\end{equation}
With this
decomposition we obtain for the correlation function Eq.~(\ref{corr}):
\begin{eqnarray}
T_\mu &=& i \epsilon_{\mu P R q'} V_{ub}
\Bigg[\frac{2 m_b g_u^{\pi\pi}(z,\zeta,W^2)}{((Pz+q')^2 - m_b^2)^2} 
\nonumber \\
&&\qquad +\frac{ f_{u\perp}^{\pi\pi}(z,\zeta,W^2)/(4W)}{(Pz+q')^2 - m_b^2}
\Bigg]
\nonumber \\
&&+\frac{V_{ub}}{(P_z + q')- m_b^2}\Bigg[ m_b P_\mu  f_{u}^{\pi\pi}(z,\zeta,W^2)
\nonumber \\&&
- (Pz+q')_\mu W e_{u}^{\pi\pi}(z,\zeta,W^2) 
\nonumber \\&&
- \frac{P_\mu R \cdot q' - R_\mu P\cdot (Pz+q')}{4W}
 f_{u\perp}^{\pi\pi}(z,\zeta,W^2)\Bigg]\;.
\end{eqnarray}
\subsection{Calculation of the decay amplitude in the LCSR approach}
\label{decc}
The next step is to relate the decay amplitude to the distribution
amplitudes described to far. For this purpose one inserts a complete 
set of states with B-meson quantum numbers between the currents 
\cite{Belyaev:1993wp,Khodjamirian:1998ji}:
\begin{eqnarray}
T_\mu(q^2)  &=&
\int d^4x e^{-iq'x}
\frac{\langle 0| \bar u(0)i \gamma_5 b(0) |B\rangle} 
    {  q^2 -m_B^2}
\nonumber \\ && \quad \times
     \langle B| \bar b(x)i \gamma_\mu(1-\gamma_5)u(x)|\pi\pi\rangle
       + \dots
\nonumber \\
&=& \frac{f_B m_B^2}{m_b(q^2- m_B^2)} {\cal M}_\mu +\dots 
\end{eqnarray}
using:
\begin{equation}
{\cal M}_\mu
=
\int d^4x e^{-iq'x} 
\langle B| \bar b(x) \gamma_\mu(1-\gamma_5)u(x)|\pi\pi\rangle\;.
\end{equation}
The ellipsis indicates all the other hadronic states which in the end
we will suppress by Borel transformation.
The next step is to define the discontinuity  Disc:
\begin{equation}
{\rm Disc}\left[ T_\mu(q^2)\right] 
= \frac{1}{2\pi i}\left( T_\mu(q^2-i\epsilon) -
T_\mu(q^2+i\epsilon)\right)\;,
\end{equation}
and isolate ${\cal M}_\mu$  using the standard duality approximation:
\begin{equation}
\int_{m_b^2}^ {s_0} ds{\rm Disc} \left[T_\mu(s)\right]
 e^{-\frac{s-m_B^2}{M^2}} \frac{m_b}{m_B^2 f_B}
= {\cal M_\mu}\;.
\end{equation}
Now one can expand ${\cal M}_\mu$ into an orthogonal-system:
\begin{eqnarray}
{\cal M}_\mu &=&
\int d^4x e^{-iq'x} 
\langle B| \bar b(x) \gamma_\mu(1-\gamma_5)u(x)|\pi\pi\rangle\;.
\nonumber \\
{\cal M}_\mu &=& { M}_1 P_\mu + { M}_2 R_\mu +
 { M}_3 q_\mu''+ \frac{i}{2W^2} { M}_4 \epsilon_{\mu PR  q''}
\nonumber \\
q''_\mu &=& q_\mu' - \frac{R\cdot q'}{R^2} R_\mu 
                   - \frac{P\cdot q'}{P^2} P_\mu\;.
\end{eqnarray}
With this decomposition one gets the following LCSR results:
\begin{eqnarray}
{ M}_1 &=& \int_0^1 \frac{dz}{z} 
\frac{ e^{-\frac{s-m_B^2}{M^2}}m_b}{m_B^2 f_B}
\Bigg[ m_b f_u^{\pi\pi}(z) - \frac{(R \cdot q')(s)}{4W} f_{u\perp}^{\pi\pi}(z)
\nonumber \\
&& \left. 
 - \frac{1}{2W}  
\left( s+(2z-1)W^2-{q'}^2 \right) 
e_u^{\pi\pi}(z)
\right]
\Theta[c(z,s_0^B)]
\nonumber \\
{ M}_2 &=& \int_0^1 \frac{dz}{z} 
\frac{ e^{-\frac{s-m_B^2}{M^2}}m_b}{m_B^2 f_B}
\Bigg[ - \frac{(R\cdot {q'})(s)}{R^2}
e_u^{\pi\pi}(z)
\nonumber \\
&& \left.
+ \frac{1}{8W}
\left( s+(2z-1)W^2-{q'}^2 \right)
f_{u \perp}^{\pi\pi}(z)
\right]
\Theta[c(z,s_0^B)]
\nonumber \\
{ M}_3 &=& -\int_0^1 \frac{dz}{z} 
\frac{ e^{-\frac{s-m_B^2}{M^2}}m_b}{m_B^2 f_B}
W e_u^{\pi\pi}(z) 
\Theta[c(z,s_0^B)]
\nonumber \\
{ M}_4 &=& \int_0^1 \frac{dz}{z} 
\frac{ e^{-\frac{s-m_B^2}{M^2}}m_b}{m_B^2 f_B}
\Bigg[
\frac{W}{2} f_{u \perp}^{\pi\pi}(z) 
\Theta[c(z,s_0^B)] 
\nonumber \\
&-&4m_bW^2 g_{u}^{\pi\pi}(z)\Bigg(
\frac{1}{zM^2} \Theta[c(z,s_0^B)] +\delta[c(z,s_0^B)]
\nonumber \\
&& \qquad - \delta[-c(z,m_b^2)]\Bigg)\Bigg]\;,
\end{eqnarray}
using:
\begin{eqnarray}
2(R\cdot q')(s) &=& (\bar \zeta-\zeta)
\left( m_B {q'}_+ -\frac{s {q'}_-}{m_B}\right)
\nonumber \\
s &=& \frac{1}{z} \left[ z\bar zW^2 + m_b^2- \bar z{q'}^2  \right]
\nonumber \\
c(z,s_0^B) &=& zs_0^B - m_b^2 + \bar z{q'}^2 - z\bar z W^2\;.
\end{eqnarray}
$\Theta$ and $\delta$ functions arise from the continuum subtraction 
 \cite{Ball:1997rj}. More precisely
we use:
\begin{eqnarray}
&&\frac{-1}{\pi}{\rm Im} \int_0^1 dz \int_{m_b^2}^{s_0^B}
\frac{ ds' e^{-\frac{s'-m_B^2}{M^2}}f(z)}{(Pz+q')^2 - m_b^2 +i\epsilon}
\nonumber \\
&=& \int_0^1\frac{dz}{z}  e^{-\frac{s-m_B^2}{M^2}}f(z)\Theta[c(z,s_0^B)]
\Theta[-c(z,m_b^2)]
\nonumber \\
&&\frac{-1}{\pi}{\rm Im} \int_0^1 dz \int_{m_b^2}^{s_0^B}
\frac{ ds' e^{-\frac{s'-m_B^2}{M^2}}f(z)}{[(Pz+q')^2 - m_b^2 +i\epsilon]^2}
\nonumber \\
&=&
-\frac{d}{d\alpha}\Bigg|_{\alpha=0}
\frac{-1}{\pi}{\rm Im} \int_0^1 dz \int_{m_b^2}^{s_0^B}
\frac{ ds' e^{-\frac{s'-m_B^2}{M^2}}f(z)}{(Pz+q')^2 - m_b^2 +i\epsilon+\alpha}
\nonumber \\
&=&-\int_0^1\frac{dz}{z}  e^{-\frac{s-m_B^2}{M^2}}f(z)
\Bigg[\frac{1}{zM^2}\Theta[c(z,s_0^B)] + \delta[c(z,s_0^B)]
\nonumber \\
&& \qquad - \delta[-c(z,m_b^2)]\Bigg]\;.
\end{eqnarray}
Note that $\Theta[-c(z,m_b^2)] = \Theta[1-z]$, and 
 $\delta[-c(z,m_b^2)] = \delta(1-z)/(m_b^2+zW^2-{q'}^2)$.
In order to obtain the square of the decay amplitude,
we have to multiply ${\cal M}_\mu {\cal M}_{\mu'}$
with the leptonic scattering tensor:
\begin{eqnarray}
L_{\mu\mu'} &=& \sum_{ss'}\bar u_s(p_e) \gamma_\mu(1-\gamma_5)
 u_{s'}(p_\nu)
\nonumber \\ && \qquad \times
  \bar u_{s'}(p_\nu) \gamma_{\mu'}(1-\gamma_5) u_s (p_e)
\nonumber \\
&=& 8\left[
(q'_\mu - p_{e\mu})p_{e\mu'} +
(q'_{\mu'} - p_{e\mu'})p_{e\mu} \right.
\nonumber \\
&& -\left.
g_{\mu\mu'}(p_e\cdot q' -m_e^2) + i\epsilon_{\mu\mu' q' p_e}\right]\;.
\end{eqnarray}
Then one gets for the matrix element:
\begin{eqnarray}
&& |M|^2=\frac{L^{\mu{\mu'}} {\cal M}_\mu {\cal M}_{\mu'}^*}
{(M_{\rm W}^2-q'^2)^2} 
\nonumber \\ & =&
\quad  \Bigg[16(M_1 P\cdot p_e + M_2 R\cdot p_e + M_3 q''\cdot p_e)
\nonumber \\ && \quad \;\times
(M_1 P\cdot q' + M_2 R\cdot q' + M_3 q''\cdot q')
\nonumber \\ && 
-16(M_1 P\cdot p_e + M_2 R\cdot p_e + M_3 q''\cdot p_e)^2
\nonumber \\ && 
+4M_4^2\frac{R^2}{W^2} 
\nonumber \\ && \times
{q''}^2 \left(
m_e^2
- \frac{(R  \cdot p_e)^2}{R^2}
- \frac{(P  \cdot p_e)^2}{W^2} 
- \frac{(q''\cdot p_e)^2}{q''^2}
\right)
\nonumber \\
&& -8 \left(
M_1^2 P^2 + M_2^2 R^2 + M_3^2 {q''}^2 - M_4^2\frac{R^2}{4W^2} {q''}^2 \right)
\nonumber \\
&& \times (p_e\cdot q' -m_e^2)
\nonumber \\
&&- 8  \frac{M_4}{W^2} \epsilon_{\mu\mu'q'p_e}\epsilon_{\mu P R q' }
\left( M_1 P_{\mu'} + M_2 R_{\mu'} + M_3 q''_{\mu'} \right) \Bigg]\;.
\nonumber \\ && \times
 \frac{1}{(M_W^2-q'^2)^2}\;.
\label{total}
\end{eqnarray}
Here we have dropped the weak coupling $(g/(2\sqrt{2}))^4$ for simplicity.
We will add it later to the phase-space element in Eq.~(\ref{result}). 
The full
expression Eq.~(\ref{total})
 is rather complicated. To simplify it we make first 
use of the fact that $P^2/{m_B'}^2$ is small, so that we can throw away all
'bad' components connected with $n^-$ which corresponds to an expansion
in $P^2/{m_B'}^2$. Then we get approximately the following expressions:

\begin{eqnarray}
2P\cdot q' &=& m_B'{q'}^+ + \frac{P^2}{m_B'} {q'}^- \to m_B{q'}^+
\nonumber \\
2p_e \cdot q' &=& (q'^2 + m_e^2)
\nonumber \\
2P\cdot p_e &=& m_B'p_{e}^+ + \frac{P^2}{m_B'}p_{e}^- \to   m_Bp_{e}^+
\nonumber \\
2 R\cdot q' &=&(\bar \zeta -\zeta) 
\left(m_B' {q'}^+ -\frac{P^2}{m_B'}{q'}^-\right)
 \to (1-2\zeta) m_B {q'}^+ \nonumber \\
2 R\cdot p_e 
 &=&(\bar \zeta -\zeta) 
\left(m_B' {p_e}^+ -\frac{P^2}{m_B'}{p_e}^-\right)
+ 4 \vec K_\perp \cdot \vec {p_{e\perp}}  
\nonumber \\
&\to& (1-2\zeta) m_B {p_e}^+
+ 4 \vec K_\perp \cdot \vec {p_{e\perp}}\;. \nonumber \\
\label{goodbad}
\end{eqnarray}
Here we have defined $p_e^+$ = $2 p_e \cdot n^+$.
The independent variables are now ${q'}^+$, $p_e^+$ and $\vec p_{e\perp}$. 
and one should note that
\begin{equation}
{\vec K_\perp}^2 = \zeta \bar \zeta W^2 - m_\pi^2\;.
\end{equation}
We can
define the dimensionless quantities: 
\begin{equation}
x = \frac{{q'}^+}{m_B}\, \quad y = \frac{p_e^+}{{q'}^+}\;.
\end{equation} 
As $m_B \gg {q'}^-$ we find also:
\begin{equation}
\frac{q'^2}{m_B^2} \ll x \;.
\end{equation}
To simplify the expression Eq.~(\ref{total}) further, we consider the limit
where q' becomes quasi light-like, i.e. $q'^2 = 0$. Then the leptonic
part of the decay process simply factorizes in analogy to the Weizs\"acker
Williams approximation in photoproduction, see e.g.~Ref.~\cite{Hoyer:2000mb}.
%
%
%
%
%
%
%\begin{eqnarray}
%&&\frac{d\Gamma_B}{d \zeta d x dy  d {q'}^2 d^2{\bf p}_{e \perp}}
%\nonumber \\
%&=&\frac{|M|^2\alpha_{\rm em}^2 |V_{ub}|^2
%}{64\sin^4(\theta_{\rm W})(4\pi)^4 m_B  y^2 x}
%\frac{\left| 1 - \frac{{q'}^2}{m_B^2 x^2}\right|}
%     {\left( 1 + \frac{p_{e \perp}^2}{m_B^2 x^2 y^2} \right)}\;.
%\label{dps}
%\end{eqnarray}
%
%
%
%
In this limit $q'$ is quasi collinear to  $p_e$,
and  one  obtains after integration over the 
angle related to ${\bf p}_{e\perp}$ and after dropping the
electron masses for the total decay amplitude:
\begin{eqnarray}
&& \frac{d\Gamma_B}{d \zeta d x dy  d {q'}^2 d p^2_{e \perp}}
\nonumber \\
&=&\frac{G_{\rm F}^2 |V_{ub}|^2}
{8(4\pi)^5 m_B   y^2 x}
\;\frac{\left(1-\frac{{q'}^2}{m_B^2 x^2}\right)}
     {\left( 1 + \frac{p_{e \perp}^2}{m_B^2 x^2 y^2}\right)}
\nonumber \\
&& \times \Bigg\{ 4 m_B^4 x^2 y(1-  y) 
\Bigg[M_1 + (1-2\zeta) M_2  
\nonumber \\ && \qquad
- M_3 \frac{m_B^2}{2 W^2}x
\left(1+ (1-2\zeta)^2\frac{W^2}{4m_\pi^2-W^2} \right)\Bigg]^2 
\nonumber \\
&& - 32 
\left( M_2- M_3 x\frac{(1-2\zeta)m_B^2}{4m_\pi-W^2}
 \right)^2 |{\bf K_\perp}|^2
| {\bf p_{e \perp}}|^2
\nonumber \\
&& \qquad + 2M_4^2 \frac{m_B^4 x^2}{W^4}
|{\bf K_\perp}|^2 |{\bf p_{e\perp}}|^2 \Bigg\}
\nonumber \\
{\rm using\;}&&  \frac{G_F}{\sqrt{2}} = 
\frac{4\pi \alpha_{\rm em}}{8 \sin^2(\theta_{\rm W}) M_W^2} \;.
\label{result}
\end{eqnarray}
\section{Discussion}
Eq.~(\ref{result}) is a simplified form of the decay width in the limits
$q'^2 \to 0$ and $P^2/{m'}^2_B \to 0$. In Appendix A we will show the
full result for finite $q'^2$ and $P^2$. Of the four distribution amplitudes
$f^{\pi\pi}$, $g^{\pi\pi}$, $e^{\pi\pi}$ and $f_\perp^{\pi\pi}$ only the asymptotic
form of  $f^{\pi\pi}$ is known and there have been attempts to model
this function and $f_\perp^{\pi\pi}$
in terms of the instanton vacuum \cite{Polyakov:1999td}.
\newline 
\newline
A calculation of the other distribution amplitudes is beyond the scope
of this article. However,
we can see if we are able at least to obtain the order of magnitude correct if
we neglect all contributions except $f^{\pi\pi}$ where we posses the expression
for the asymptotic form  and
compare  the semi-leptonic decay $B\to \pi\pi\ell\nu$ with
the semi-leptonic decay  $B\to \rho\ell\nu$. 
Neglecting $e^{\pi\pi}$, $h^{\pi\pi}$ and $g^{\pi\pi}$ may
not be so unreasonable, as the first one is twist-3 (i.e. a higher-twist contribution)
and the other two are connected to 'polarization' states where we know from the
experiences of spin physics that the contribution of the quarks is relatively small.
If, in this sense, we only retain the contribution
from $f^{\pi\pi}$, we obtain a simple expression for
for the branching ratio  $B(B\to \pi\pi\ell\nu)$:
\begin{eqnarray}
 \frac{d B(B\to \pi\pi\ell\nu)}{d \zeta d x dy  d {q'}^2 d p^2_{e \perp}}
&=&\frac{G_{\rm F}^2|V_{ub}|^2}
{2(4\pi)^5 m_B \Gamma_{B}  y^2 x}
\;\frac{\left(1-\frac{{q'}^2}{m_B^2 x^2}\right)}
     {\left( 1 + \frac{p_{e \perp}^2}{m_B^2 x^2 y^2}\right)}
\nonumber \\
&& \times   m_B^4 x^2 y(1-  y) \left(  M_1^{\rm twist-2}\right)^2
\nonumber 
\end{eqnarray}
\begin{equation}
M_1^{\rm twist-2}
 = \int_0^1 \frac{dz}{z} e^{-\frac{s-m_B^2}{M^2}}
\Theta[c(z,s_0^B)]
\frac{m_b^2}{m_B^2 f_B} f_u^{\pi\pi}(z)\;, 
\nonumber \\
\label{practicalresult}
\end{equation}
where $\Gamma_B$ is the total B-meson decay width.
\section{Numerical estimates}
\label{numerics}
\begin{figure}
\centerline{\psfig{figure=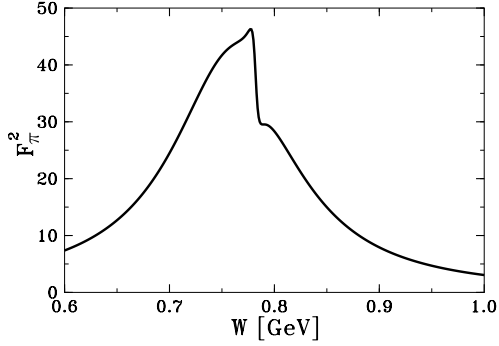,width=8.0cm}}
\caption{Square of the time-like pion form factor ($F_\pi^2$)
in the HLS parameterization.}
\label{timefpi}
\end{figure}

\begin{figure}
\centerline{\psfig{figure=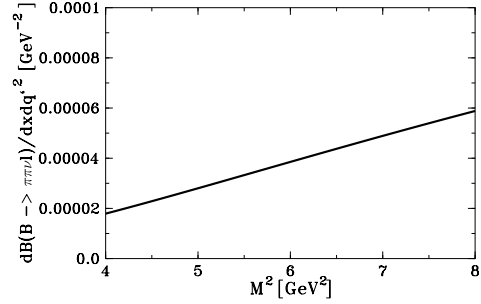,width=8.0cm}}
\caption{Borel dependency of the differential branching ratio
$B\to \pi\pi\ell\nu$. The value for $x$ is fixed at the $\rho$-pole
in the HLS parameterization, i.e.~$x = 1-m^2_{\rho, \rm HLS}/m_B^2$, 
with $m_{\rho, \rm HLS} = 774.57 \;{\rm MeV}$.}
\label{borel}
\end{figure}

\begin{figure}
\centerline{\psfig{figure=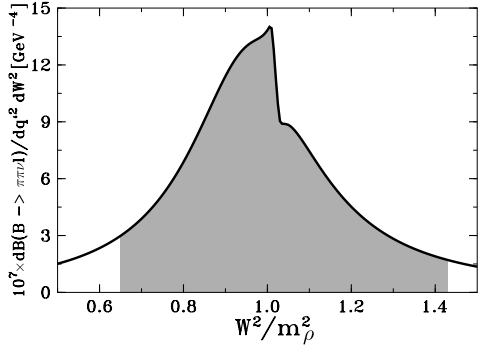,width=8.0cm}}
\caption{Integration range over the $\rho$ pole: The shaded
area depicts the range of integration in $W^2$ of the differential
branching ratio $B\to \pi\pi\ell\nu$.}
\label{w}
\end{figure}

Now, we try to make some order of magnitude estimates for a comparison
between the semi-leptonic decay of a B meson 
into two pions on the one hand and into a $\rho$ meson on the other hand.
For $f^{\pi\pi}$  one can use the asymptotic form given in
Ref.~\cite{Polyakov:1999td}:
\begin{equation}
f_u^{\pi\pi\;{\rm as}}(z,\zeta,W^2) = 6 z(1-z) (2\zeta-1) F_\pi(W^2)\;.
\end{equation}
$F_\pi(W^2)$ is the pion form factor in the time-like region, normalized
by $F_\pi(0) =1$. 
For the pion form factor in the time-like
region we use the fit from the CMD2-Collaboration \cite{Akhmetshin:1999uj} 
using the Hidden Local Symmetry (HLS) parameterization, which
is displayed in  Fig.~\ref{timefpi}. The shape of the time-like pion
form factor is a characteristic superposition of the $\omega$ and $\rho$ 
resonances.
\newline \newline
For the numerics we take 
$G_{\rm F} = 1.16639 \times 10^{-5}\;{\rm GeV}^{-2}$ ,
$\Gamma_B^{-1}=1.62\times 10^{-12}\;{\rm s}$,
$m_B= 5.279 \;{\rm GeV}$, and $V_{\rm ub}=
0.0035$ \cite{Groom:2000in}. The value of $V_{\rm ub}$ is 
an average value where the error assigned to it is of the
order of  50\%.
For the decay width 
$f_B$ we make use of the corresponding 
sum rule expression, see e.g.~Ref.~\cite{Ali:1994vd}:
\begin{eqnarray}
f_B^2 &=& \frac{m_b^2}{m_B^4}\exp\left(\frac{m_B^2-m_b^2}{M^2}\right) \Bigg[
-m_b \langle \bar q q \rangle_{\mu^2=M^2} \nonumber \\
&& - 
\frac{m_b}{2M^2} \left(1-\frac{m_b^2}{2M^2}\right)\langle \bar q \sigma g G q \rangle_{\mu^2=M^2} 
\nonumber \\
&&
+\frac{3}{8\pi^2} \int_{m_b^2}^{s_0^B} sds e^{-\frac{s-m_b^2}{M^2}}
\left(1-\frac{m_b^2}{s}\right)^2 \Bigg]
\end{eqnarray}
\begin{eqnarray}
 \langle \bar q q \rangle_{\mu^2} &=& 
\left(\frac{\alpha_s(\mu)}{\alpha_s(\mu_0)}\right)^{-4/\beta_0}
 \langle \bar q q \rangle_{\mu_0^2}  
\nonumber \\
\langle \bar q \sigma g G q \rangle_{\mu^2} &=&
\left(\frac{\alpha_s(\mu)}{\alpha_s(\mu_0)}\right)^{2/(3\beta_0)}
\langle \bar q \sigma g G q \rangle_{\mu_0^2} 
\nonumber \\
\beta_0 &=& \frac{11}{3}N_c - \frac{2}{3}n_f
\nonumber \\
 \langle \bar q q \rangle_{ 1\;{\rm GeV}^2} &=& -245\;{\rm MeV}^3
\nonumber \\
\langle \bar q \sigma g G q \rangle_{1\;{\rm GeV}^2} &=& 
0.65\;{\rm GeV}^2  \langle \bar q q \rangle_{ 1\;{\rm GeV}^2}\;.
\end{eqnarray}
The values of the condensates at $\mu_0$ = 1 GeV have been taken from \cite{Ball:1997rj}.
In the formula for $f_B$ radiative corrections are not taken into account because they
are  not taken into account in all the other LCSR calculations presented or used
here either. For the same reason we use for $\alpha_s$ the one-loop expression:
\begin{equation}
\alpha_s(\mu) = \frac{4\pi}{\beta_0 \ln(\mu^2/\Lambda^2)}\;,
\end{equation}
using $\Lambda^{(5)} = 0.208\;{\rm GeV}$ and for the threshold masses
in the evolution $m_c$ = 1.25 GeV and $m_b$ = 4.2 GeV \cite{Groom:2000in}.
The sum rule parameters are chosen as in \cite{Ball:1997rj}. They
can be obtained from Tab.~\ref{srparams}
\begin{table}
\begin{tabular}{|c||c|c|c|}
\hline 
&&&\\
& $M^2$ [GeV]$^2$ & $s_0^B$ [GeV]$^2$  & $m_b$ [GeV] \\
&&&\\
\hline 
&&&\\
left border   & 4 &  35 & 4.7 \\ 
&&&\\
central value & 6 &  34 & 4.8 \\
&&&\\
right border  & 8 &  33 & 4.9 \\
&&&\\
\hline
\end{tabular}
\caption{Sum rule parameters used in the calculation
for the estimate of the B-decay width. For the calculation
of the decay width the central values of the sum rule parameters 
have been used, and
for the error the maximal deviation to the left and right border.}
\label{srparams}
\end{table}
%
%
%
%$4\;{\rm  GeV}^2 < M^2 < 8 \;{\rm  GeV}^2$,
%$35\;{\rm  GeV}^2 > s_0^B > 33 \;{\rm  GeV}^2$, and
%$4.7\;{\rm  GeV}^2 < m_b < 4.9 \;{\rm  GeV}$. For the calculations
For the calculation
of the decay width the central values have been used, and
for the error the maximal deviation to the left and right border.
For the dependency on the Borel parameter  in 
Fig.~\ref{borel} we consider the differential
branching ratio  at ${q'^2} =0$ using the central values
for the other sum rule parameters and integrating
 $p_{e\perp}^2 \in [0,m_B^2x^2/4]$, $y \in [0,1]$,
$\zeta\in [0,1]$. The value for $x$ is fixed at the $\rho$-pole
in the HLS parameterization, i.e.~$x = 1-m^2_{\rho, \rm HLS}/m_B^2$, 
with $m_{\rho, \rm HLS} = 774.57 \;{\rm MeV}$. This means
$x \approx 0.97847$, so effectively $x$ is close to 1 if the 
invariant mass of the two pions is in the vicinity of the $\rho$
meson pole. 
It can be seen that the dependency on the
Borel parameter is rather strong, as we only consider the asymptotic
form without any higher-twist contributions or radiative corrections.
Furthermore, for simplicity, we have kept the other sum rule parameters
fixed. In principle they should vary with the Borel parameter as 
given e.g.~in Tab.~\ref{srparams}.
\newline \newline
As the $\rho$ meson decays
nearly exclusively into two pions there should be a chance to match
the branching ratio for the semi-leptonic decay $B\to \pi\pi\ell\nu$ with the
corresponding decay $B\to \rho\ell\nu$. More precisely we have to integrate
the branching ratio over $W\in[m_\rho-\Gamma_\rho, m_\rho+\Gamma_\rho]$, 
i.e.~the $\rho$-meson pole and compare:
\begin{equation}
\frac{dB(B^\pm \to \nu \ell^\pm  \pi^+\pi^-)}{d{q'}^2}\Bigg|
_{{q'}^2=0} \equiv 
\frac{dB(B^0 \to \nu e^- \rho^+)}
{4 d{q'}^2}\Bigg|_{{q'}^2=0}\;.
\label{comparison}
\end{equation}
Here one factor 1/2 accounts for the fact that we have to consider
a $\rho^0$ wave function instead for a $\rho^+$  and another factor 1/2
takes into account that we only consider charged pions. 
$B(\bar B^0 \to \rho^+e^-\bar \nu)$ at ${q'}^2=0$ only depends
on two form factors $A_1(0)$ and $A_2(0)$ \cite{Ball:1997rj}:
\begin{eqnarray}
&&\frac{dB(\bar B^0 \to \rho^+e^-\bar \nu)}{d{q'}^2}\Bigg|_{{q'}^2=0}
= \frac{G_{\rm F}^2 |V_{\rm ub}|^2}{192\pi^3 m_B^3\Gamma_B}
\sqrt{\lambda}H_0^2
\nonumber \\
&&\lambda = (m_B^2 + m_\rho^2)^2 -4m_B^2m_\rho^2
\nonumber \\
&&H_0 = \frac{1}{2m_\rho}
\Bigg[(m_B^2 - m_\rho^2)(m_B+m_\rho) A_1(0)
\nonumber \\ && \quad
- \frac{\lambda}{m_B+m_\rho} A_{2}(0)  \Bigg]\;.
\label{brho}
\end{eqnarray}
For the two form factors $A_1(0)$ and $A_2(0)$ we use the values
in Ref.~\cite{Ball:1997rj}:
\begin{eqnarray}
A_1(0) &=& 0.27\pm 0.05
\nonumber \\
A_2(0) &=& 0.28\pm 0.05\;.
\end{eqnarray}
For the integration of the branching ratio $B(B\to \pi\pi\ell\nu)$ over the 
$\rho$ pole we take the range 
$W^2 \in [ (m_\rho-\Gamma_\rho)^2,(m_\rho+\Gamma_\rho)^2]$, see  Fig.~\ref{w}.
Hereby we use for the variable transformation from $x$ to $W^2$ the fact that
\begin{equation}
W^2 = P^2 = m_B^2 \frac{1-x}{x}\left(x-\frac{{q'}^2}{m_B^2}\right) \to
m_B^2 (1-x)\;.
\end{equation}
We can now compare the the semi-leptonic branching ratio
 $B\to \pi\pi\ell \nu$ with the  equivalent $B\to \rho\ell\nu$ where the additional
factors are taken into account to make the two quantities comparable.
Using the central values of the sum rule  parameters for the calculation
and obtaining the error from varying the sum rule parameters over the allowed
range we obtain:
\begin{eqnarray}
\frac{dB(\bar B^0\to\rho^+\ell^-\bar \nu)}{4d {q'}^2} 
\Bigg|_{{q'}^2=0} 
&=& \left(13.8^{+ 28.1}_{-13.8}\right) \times 10^{-7}\;{\rm GeV}^{-2}
\nonumber \\
\frac{dB(B\to\pi\pi\ell\nu)}{d {q'}^2} \Bigg|_{{q'}^2=0} 
&=& (3.0 \pm 2.7) \times 10^{-7}\;{\rm GeV}^{-2}\;.
\nonumber \\
\end{eqnarray}
The consistency in the order of magnitude 
can be taken as a hint that our approach is qualitatively
correct. In our case the large error results from the fact that
we did neither take into account all 2$\pi$DAs nor the higher-twist 
corrections  and restricted
ourselves to the asymptotic form only. The big error in case of the
$B\to \rho$ semi-leptonic decay comes from the fact that in the kinematic
region we consider here, the decay width is 
the the result of two contributions that nearly cancel each other, 
c.~f.~Eq.~(\ref{brho}). The number comes from the usual error-analysis,
and reveilles drastic effects. We should state that this is not a numerical
estimate of the decay width, but rather a consistency check, that we get
the order of magnitude correct. The important task that remains to be done
is a modeling of the 2$\pi$DAs which will then allow for a quantitative
prediction of the semi-leptonic decay of B mesons into two pions.
\section{Summary and conclusions}
To summarize, we have shown that the semi-leptonic decay 
$B^\pm \to \pi^+\pi^-\nu \ell^\pm$ can be described in the LCSR formalism
using the two-pion distribution amplitudes. Major observables, except for
$f^{\pi\pi}$ and $f_\perp^{\pi\pi}$, are the twist-2
distribution amplitude  $g^{\pi\pi}$ and  the twist-3 amplitude $e^{\pi\pi}$. 
When we retain  only the twist-2 distribution amplitude $f^{\pi\pi}$, where
the asymptotic form is known, 
and compare the semi-leptonic decay width $B\to \pi\pi\ell\nu$ with the
corresponding decay width $B\to\rho\ell\nu$ at ${q'}^2=0$, we find consistency
in the order of magnitude.
\newline \newline
\indent
I wish to acknowledge  A.~Khodjamirian and C.~Weiss for fruitful discussion.

\begin{appendix}
\label{app}
\section{The full formula for the decay width}
In the following we give the formula for the decay width
$B^+ \to  \pi^+\pi^-\ell^+\nu_\ell$ for finite ${q'}^2$ and $P^2$.
The electron mass is neglected in the calculation altogether.
The formula can be obtained from Eq.~(\ref{total}) by the following
substitutions:
\begin{eqnarray}
2 P\cdot q' &=&  m_B^2 x + \frac{x-\bar x}{x} {q'}^2
\nonumber \\
2p_e \cdot q' &=& q'^2
\nonumber \\
2P\cdot p_e &=&  (m_B^2 x - {q'}^2)y + \frac{\bar x}{x}
\frac{p_{e\perp}^2}{y}
\nonumber \\
2 R\cdot q' &=& \frac{\bar \zeta -\zeta}{x}[m_B^2 x^2 + {q'}^2]
\nonumber \\
2 R\cdot p_e &=& (\bar \zeta -\zeta)
\left[(m_B^2 x - {q'}^2)y 
- \frac{\bar x}{x} \frac{p_{e\perp}^2}{y} \right] 
\nonumber\\ && \qquad \qquad 
+ 4 \vec K_\perp\cdot \vec p_{e\perp}
\nonumber \\ &=&
(2 R\cdot p_e)_\parallel + 4 \vec K_\perp
\cdot \vec p_{e\perp}\;.
\end{eqnarray}
In the limit ${q'}^2\to 0$ and $\bar x \to 0$ we reproduce the formulas
Eq.~(\ref{goodbad}). For completeness we add here once more the expressions
for $R^2$, $P^2$, and $|\vec K_\perp|^2$:
\begin{eqnarray}
P^2 &=& W^2 = m_B^2 \frac{1-x}{x}\left(x-\frac{{q'}^2}{m_B^2}\right) 
\nonumber \\
R^2 &=& 4 m_\pi^2 -  W^2
\nonumber \\
{\vec K_\perp}^2 &=& \zeta\bar \zeta W^2 - m_\pi^2\;.
\end{eqnarray}
Using the expressions above we can write down
the total decay width, integrated over the polar angle of 
$\vec p_{e\perp}$:
\begin{eqnarray}
&& \frac{d\Gamma_B}{d \zeta d x dy  d {q'}^2 d p^2_{e \perp}}
\nonumber \\
&=&\frac{G_{\rm F}^2 |V_{ub}|^2}
{8(4\pi)^5 m_B   y^2 x}
\;\frac{\left(1-\frac{{q'}^2}{m_B^2 x^2}\right)}
     {\left( 1 + \frac{p_{e \perp}^2}{m_B^2 x^2 y^2}\right)}
\nonumber \\
&& \times \Bigg\{16
\Bigg[M_1 P\cdot p_e + M_2 (R\cdot p_e)_\parallel 
\nonumber \\
&& + 
M_3 \left(q'\cdot p_e
-\frac{(R\cdot q')( R\cdot p_e)_\parallel}{R^2}
-\frac{(P\cdot q')( P\cdot p_e)}{P^2}\right)\Bigg]
\nonumber \\ && \quad \;\times
\Bigg[M_1 P\cdot q' + M_2 R\cdot q' 
\nonumber \\
&& + 
M_3 \left({q'}^2
-\frac{(R\cdot q')^2}{R^2}
-\frac{(P\cdot q')^2}{P^2}\right)\Bigg]
\nonumber \\ && 
\nonumber \\ && 
\nonumber \\ && 
-16\Bigg[M_1 P\cdot p_e + M_2 (R\cdot p_e)_\parallel 
\nonumber \\ && +
M_3 \left(q'\cdot p_e
-\frac{(R\cdot q')( R\cdot p_e)_\parallel}{R^2}
-\frac{(P\cdot q')( P\cdot p_e)}{P^2}\right)\Bigg]^2
\nonumber \\ && 
-32 |\vec K\perp|^2 |\vec p_{e\perp}|^2 \left( M_2 - M_3\frac{R\cdot q'}{R^2}
\right)^2
\nonumber \\ 
\nonumber \\  
\nonumber \\ &&  
+ 4 M_4^2 \frac{R^2}{P^2}
\Bigg[-\left( {q'}^2 -\frac{(R\cdot q')^2}{R^2} 
                    - \frac{(P\cdot q')^2}{P^2}
\right)
\nonumber \\ && \qquad\qquad \times
\left(         \frac{(R\cdot p_e)^2_\parallel}{R^2}
                    + \frac{(P\cdot p_e)^2}{P^2}
\right)
\nonumber \\
&& \qquad - \left( p_e q' - \frac{(R\cdot p_e)_\parallel(R\cdot q')}{R^2}
                   - \frac{(P\cdot p_e)(P\cdot q')}{P^2}
\right)^2
\nonumber \\  &&\qquad \qquad 
- 2\frac{|\vec K_\perp|^2|\vec p_{e\perp}|^2}{R^2}
\left({q'}^2 - \frac{(P\cdot q')^2}{P^2} \right)
\Bigg]
\nonumber \\
&& -4q'^2 \Bigg[
M_1^2 P^2 + M_2^2 R^2 + 
\nonumber \\ && \quad + \left(M_3^2  - M_4^2\frac{R^2}{4P^2} \right)
\left({q'}^2 - \frac{(P\cdot q')^2}{P^2}- \frac{(R\cdot q')^2}{R^2}\right)
\Bigg]
\nonumber \\
\nonumber \\
\nonumber \\
&&- 8  \frac{M_4}{P^2} \Bigg[
\Bigg(P\cdot q' \left[(P\cdot p_e)(R\cdot q')  -(P\cdot q')(R\cdot p_e)_\parallel\right]
\nonumber \\ && \qquad 
+P^2 {q'}^2 \left[ (R\cdot p_e)_\parallel - \frac{1}{2} R\cdot q'  \right] 
\Bigg)
\Bigg( M_1 - M_3\frac{P\cdot q'}{P^2}\Bigg)
\nonumber \\ && \qquad \quad -
\Bigg(R\cdot q' \left[(R\cdot p_e)_\parallel(P\cdot q')  -(R\cdot q')(P\cdot p_e)\right]
\nonumber \\ && \qquad 
+R^2 {q'}^2 \left[ P\cdot p_e - \frac{1}{2} P\cdot q'  \right] 
\Bigg)
\Bigg( M_2 - M_3\frac{R\cdot q'}{R^2}\Bigg)\Bigg]\Bigg\}
\nonumber \\ && \times
 \frac{M_W^4}{(M_W^2-q'^2)^2}\;.
\label{full}
\end{eqnarray}
In the form given above the decay width is easy to program in a computer
code.
\end{appendix}
\end{document}